# Polarization based intensity correlation of polarization speckle

ABHIJIT ROY AND MARUTHI M. BRUNDAVANAM

*Department of Physics, Indian Institute of Technology Kharagpur, Kharagpur, India, 721302*
*E-mail: abhijitphy302@gmail.com, bmmanoj@phy.iitkgp.ac.in*

**A new, different kind of intensity correlation, denoted as polarization based intensity correlation, is proposed and investigated to study the correlation between different polarization components of polarization speckle, which has non-uniform spatial polarization distribution. It is shown both theoretically and experimentally that the range of the polarization based intensity correlation for a particular polarization component of the polarization speckle depends on the spatial average intensity of the speckles corresponding to that particular polarization component. The experimentally determined nature of the change of range of the intensity correlation for different polarization components, due to variation in the average intensity, is found to be matching well with the theoretical prediction. The existence of non-zero correlation between two orthogonally polarized speckle patterns, filtered from a partially depolarized speckle pattern, is also observed. This study may be useful in exploiting the polarization based intensity correlation for different applications such as speckle cryptography etc.**

A randomly distributed granular intensity pattern, known as speckle pattern, is observed when a rough surface or a diffusive medium is illuminated with a coherent beam of light [1]. Although the intensity is randomly distributed in a speckle pattern, it has been observed that the information of the input beam is scrambled into it, and different techniques have been developed to retrieve the information from a speckle pattern [1, 2]. The introduction of the intensity correlation to determine the angular correlation of speckles has paved the way to uncover an important property of the speckles: the memory effect [3]. Subsequently, different types of memory effects, such as, the translational memory effect and rotational memory effect, based on the intensity correlation, have also been proposed and demonstrated [4, 5]. Different techniques exploiting the two-point intensity correlation of a speckle pattern have been demonstrated for non-invasive imaging through a scattering medium [6, 7]. Higher order intensity correlation based approach is also reported to be useful in retrieving the phase information from a recorded intensity distribution, which is useful for looking through barrier [8]. The average size of speckle grains in a speckle pattern, which is required for proper characterization of the speckle pattern, is also determined following the two-point intensity correlation technique [9].

Apart from the aforementioned applications, the intensity correlation technique is also reported to be a very useful tool to investigate the spatial coherence-polarization property, another important characteristics, of a speckle pattern. The spatial coherence-polarization property of a speckle pattern is characterized from the study of the spatial degree of coherence (DoC) and degree of polarization (DoP), which are determined from the two-point intensity correlation function of the speckle pattern [10, 11]. The speckles, based on their spatial coherence-polarization property, are categorized into intensity speckle (with uniform spatial polarization distribution) and polarization speckle (with non-uniform spatial polarization distribution). Although the polarization speckle has tremendous potential application in various branches of science, not much attention have been paid to characterize the polarization speckle properly, until recently. Different techniques based on the interferometric approach have been proposed to generate and control the depolarization of the speckles [10, 12]. Separately, polarization speckle has also been reported to be generated using a bulk scattering medium or a spiral phase plate [13, 14]. Recently, an analysis of polarization speckle has been reported for imaging of polarized objects through a birefringent scattering medium [15]. In another study, polarization memory effect of polarization speckle, which is generated from a birefringent scatterer, has been investigated [16].

Although different studies have been reported on the generation and analysis of polarization speckle, a detailed investigation on the characterization of polarization speckle is not reported yet. A recently introduced technique, which combines the intensity cross-correlation approach with polarization, has been reported to be useful to track the polarization rotation and also to retrieve the Stokes parameters of an input beam through a bulk scattering medium [17, 18]. In this work, the intensity cross-correlation technique combined with the polarization, which is referred as polarization based intensity correlation in this paper, is employed to study the correlation between different polarization components of polarization speckle, which is generated from a multiple scattering medium. A detailed theoretical analysis along with the experimental demonstration are presented.

Let us consider a transverse spatially random monochromatic electric field vector, $\mathbf{E}(\mathbf{r}, t)$ is composed of two mutually orthogonal polarization components $E_x(\mathbf{r}, t)$ and $E_y(\mathbf{r}, t)$, and is written as

$$\mathbf{E}(\mathbf{r}, t) = E_x(\mathbf{r}, t)\,\hat{x} + E_y(\mathbf{r}, t)\,\hat{y}, \quad (1)$$

where $\hat{x}$ and $\hat{y}$ are the two unit orthogonal polarization vectors, $\mathbf{r}$ is the spatial position vector on the transverse observation plane, and t is the time. A spatially random field is characterized from the study of the spatial degree of coherence, $\gamma(\mathbf{r}_1, \mathbf{r}_2)$ and the spatial degree of polarization, $P(\mathbf{r})$ of the random field, following the second order or two-point intensity correlation technique as [1]

$$\gamma^2(\mathbf{r}_1, \mathbf{r}_2) = \frac{\langle \Delta I(\mathbf{r}_1)\,\Delta I(\mathbf{r}_2)\rangle}{\langle I(\mathbf{r}_1)\rangle\,\langle I(\mathbf{r}_2)\rangle}, \quad (2)$$

$$P^2(\mathbf{r}) = 2\gamma^2(\mathbf{r}, \mathbf{r}) - 1, \quad (3)$$

where $\Delta I(\mathbf{r}) = I(\mathbf{r}) - \langle I(\mathbf{r})\rangle$ is the fluctuation of intensity from its mean value, and '<>' denotes the ensemble average of the variable. The random field, $\mathbf{E}(\mathbf{r})$ after passing through a polarizer with its pass axis oriented at an angle θ with respect to the x-axis is modified as

$$\mathbf{E}_P(\mathbf{r}) = [\cos^2\theta\,E_x(\mathbf{r}) + \sin\theta\cos\theta\,E_y(\mathbf{r})]\,\hat{x} + [\sin\theta\cos\theta\,E_x(\mathbf{r}) + \sin^2\theta\,E_y(\mathbf{r})]\,\hat{y}. \quad (4)$$

The polarization based intensity correlation of polarization speckle is investigated by studying the two-point intensity cross-correlation of the speckle patterns, which are recorded at different θ of the polarizer. The square of the degree of cross-correlation, $\gamma_c(\mathbf{r}_1, \mathbf{r}_2)$ of two speckle patterns is defined as [1]

$$\gamma_c^2(\mathbf{r}_1, \mathbf{r}_2) = \frac{\langle \Delta I(\mathbf{r}_1)\,\Delta I(\mathbf{r}_2)\rangle}{\sigma_I(\mathbf{r}_1)\,\sigma_I(\mathbf{r}_2)}, \quad (5)$$

where $\mathbf{r}_1$ and $\mathbf{r}_2$ are two spatial position vectors on the transverse plane of two speckle patterns, and $\sigma_I$ is the standard deviation of intensity. The intensity, I and the average intensity, $\langle I(\mathbf{r})\rangle$ of the random field, $\mathbf{E}_P(\mathbf{r})$ can be written as

$$I = \cos^2\theta\,E_x\,E_x^* + \sin\theta\cos\theta\,E_x^*\,E_y + \sin\theta\cos\theta\,E_x\,E_y^* + \sin^2\theta\,E_y\,E_y^*. \quad (6)$$

$$\langle I\rangle = \cos^2\theta\,\Gamma_{xx} + \sin\theta\cos\theta\,\Gamma_{xy} + \sin\theta\cos\theta\,\Gamma_{yx} + \sin^2\theta\,\Gamma_{yy}, \quad (7)$$

where $\Gamma_{ij}$s are the elements of coherence matrix of the random field $\mathbf{E}(\mathbf{r}, t)$, and are defined as $\Gamma_{ij}(\mathbf{r}_1, \mathbf{r}_2) = \langle E_i^*(\mathbf{r}_1)\,E_j(\mathbf{r}_2)\rangle$. The $\gamma_c^2(\mathbf{r}_1, \mathbf{r}_2)$ of two speckle patterns, recorded for the pass axis of the polarizer at $\theta_1$ and $\theta_2$, is calculated by inserting Eqs. 6 and 7 into Eq. 5, under the assumptions that the speckle field follows the Gaussian statistics, and the speckles are spatially stationary and ergodic in nature, and is found to be

$$\gamma_c^2(\mathbf{r}_1, \mathbf{r}_2) = \frac{\langle \Delta I(\mathbf{r}_1)\,\Delta I(\mathbf{r}_2)\rangle}{\sigma_I(\mathbf{r}_1)\,\sigma_I(\mathbf{r}_2)} = \frac{|N|^2}{D_1 \times D_2}, \quad (8)$$

where $N = \cos\theta_1\cos\theta_2\,\Gamma_{xx} + \cos\theta_1\sin\theta_2\,\Gamma_{xy} + \sin\theta_1\cos\theta_2\,\Gamma_{yx} + \sin\theta_1\sin\theta_2\,\Gamma_{yy}$ and $D_t = \cos^2\theta_t\,\Gamma_{xx} + \sin\theta_t\cos\theta_t\,\Gamma_{xy} + \sin\theta_t\cos\theta_t\,\Gamma_{yx} + \sin^2\theta_t\,\Gamma_{yy}$. Following the conditions of complete depolarization in the field i.e. $\Gamma_{ij}(\mathbf{r}, \mathbf{r}) = \langle E_i^*(\mathbf{r})\,E_j(\mathbf{r})\rangle = 0$ for $i \neq j$ and $\langle |E_x(\mathbf{r})|^2\rangle = \langle |E_y(\mathbf{r})|^2\rangle$, the values of N and $D_t$ at $\mathbf{r}_1 = \mathbf{r}_2$ can be found as $\cos(\theta_1 - \theta_2)$ and 1, respectively, and inserting these values in Eq. 8, the $\gamma_c^2(\mathbf{r}, \mathbf{r})$ can be calculated as

$$\gamma_c^2(\mathbf{r}, \mathbf{r}) = |\cos(\theta_1 - \theta_2)|^2. \quad (9)$$

In order to study the polarization based intensity correlation of a particular polarization, the corresponding polarization orientation, here $\theta_1$ in Eq. 9, is considered as reference, and the speckles recorded at $\theta_1$ are cross-correlated with the speckles captured at other orientations of the pass axis of the polarizer, here $\theta_2$ in Eq. 9, where $\theta_2$ is varied from $\theta_1 - 90^0$ to $\theta_1 + 90^0$. It can be observed from Eq. 9 that the maximum $\gamma_c^2$ changes sinusoidally from zero to unity as a function of the difference between the orientations of the polarization vectors of the two cross-correlating speckle patterns. The full width at half maxima (FWHM) of $\gamma_c^2(\mathbf{r}, \mathbf{r})$ for a particular $\theta_1$ is considered as the range of the polarization based intensity correlation of that particular polarization component. It can also be found from Eq. 9 that for a completely depolarized random field, the range of the polarization based intensity correlation is same for all the polarization components of the field i.e. the range is independent of the choice of $\theta_1$.

If the average intensities of the horizontal and vertical polarization components of the random field are not same and the intensity ratio is defined as $\alpha = \frac{\Gamma_{xx}(\mathbf{r},\mathbf{r})}{\Gamma_{yy}(\mathbf{r},\mathbf{r})} = \frac{\langle I_x\rangle}{\langle I_y\rangle}$, the numerator and denominator in Eq. 8 are modified as $N = \Gamma_{yy}(\alpha\cos\theta_1\cos\theta_2 + \sin\theta_1\sin\theta_2)$ and $D_t = \Gamma_{yy}(\alpha\cos^2\theta_t + \sin^2\theta_t)$, respectively. After inserting these values, Eq. 8 is modified as

$$\gamma_c^2(\mathbf{r}, \mathbf{r}) = \frac{|\alpha\cos\theta_1\cos\theta_2 + \sin\theta_1\sin\theta_2|^2}{(\alpha\cos^2\theta_1 + \sin^2\theta_1)(\alpha\cos^2\theta_2 + \sin^2\theta_2)}. \quad (10)$$

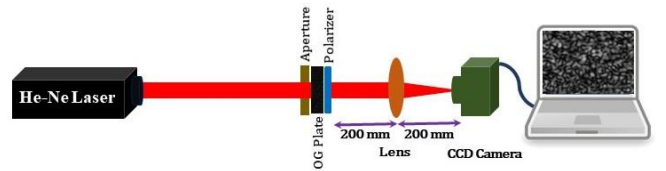

Fig. 1. The schematic diagram of the experimental setup.

The schematic diagram of the experimental setup is shown in Fig. 1. A random bulk scattering medium, here an opal glass (OG) plate, is illuminated by a horizontally polarized light of wavelength of 632.8 nm originating from a He-Ne laser source. The beam size on the OG plate is adjusted using an aperture, and a beam of diameter of 3 mm is passed through the scattering medium. The multiple scattering encountered by the coherent beam of light, while propagating through the OG plate, makes the generated speckle pattern spatially depolarized, and these types of speckles are known as polarization speckle. As shown in Fig. 1, a Fourier arrangement is constructed using a bi-convex lens having focal length of 200 mm, and by placing the OG plate at the front focal plane and a CCD camera at the rear focal plane of the Fourier transforming lens. The Fourier arrangement is made to record the far-field speckle pattern and to achieve spatial stationarity in the recorded speckle pattern. In order to investigate the polarization based intensity correlation of the polarization speckle, a polarizer is placed after the OG plate, and the filtered speckle patterns for different orientations (θ) of the pass axis of the polarizer (from $0^0$ to $360^0$ in step of $10^0$) are recorded. To characterize the speckles generated from the OG plate, the polarizer is removed and the far-field speckles are recorded.

The speckle patterns recorded with the polarizer at θ = $0^0$ and $90^0$, and without the polarizer are characterized following the intensity correlation technique presented in Eq. 2, where the ensemble average is replaced by the spatial average under the assumptions of spatial ergodicity and stationarity of the recorded speckle patterns, and the results are presented in Fig. 2. It can be observed from Fig. 2 that the squares of the maximum degrees of coherence in the presence of the polarizer are very close to 1.0, as expected in case of a fully spatially polarized speckle pattern, which can also be concluded from Eq. 3. On the other hand, the square of the maximum degree of coherence in the absence of the polarizer is found to be 0.5, which confirms from Eq. 3 that the speckles generated from the OG plate are spatially depolarized.

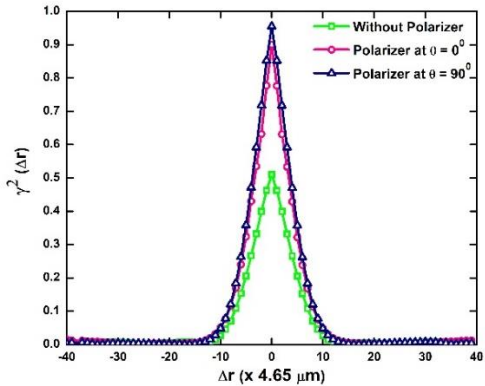

Fig. 2. The far-field intensity correlation functions of speckle patterns in the presence and absence of the polarizer.

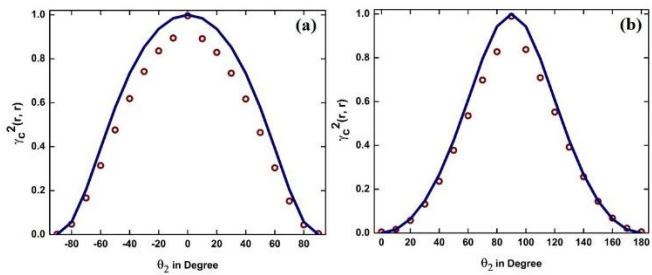

Fig. 3. The experimentally observed (circles) and theoretically predicted (solid line) variations of the maximum $\gamma_c^2$ with $\theta_2$ for: $\theta_1$ = (a) $0^0$ and (b) $90^0$.

The polarization based intensity correlation of the depolarized speckles is investigated from the study of intensity cross-correlation between the speckle patterns, which are recorded at different orientations of the pass axis of the polarizer. The intensity cross-correlation between two speckle patterns is studied following Eq. 5. As discussed earlier, in order to study the polarization based intensity correlation of a particular polarization component with polarization orientation $\theta_1$, the change of the maximum $\gamma_c^2$ with the polarization orientation ($\theta_2$) of the cross-correlating speckle patterns, which ranges from $\theta_2 = \theta_1 + 90^0$ to $\theta_2 = \theta_1 - 90^0$, is investigated. The experimentally observed variations of the maximum $\gamma_c^2$ with $\theta_2$, in case of $\theta_1 = 0^0$ and $90^0$, are presented in circles in Figs. 3(a) and 3(b), respectively. It can be observed that the FWHM of the variation of the maximum $\gamma_c^2$ with $\theta_2$, refereed as the range of the intensity correlation, are not same in these two cases, and are found to be $112.12^0$ and $71.48^0$, respectively. In order to study the existence of any systematic change in the range of the intensity correlation with $\theta_1$, the range of the intensity correlation of different polarization components of the depolarized field i.e. for different $\theta_1$ is determined, and is found to be changing monotonically with $\theta_1$, and the variation is presented as circles in Fig. 4(a).

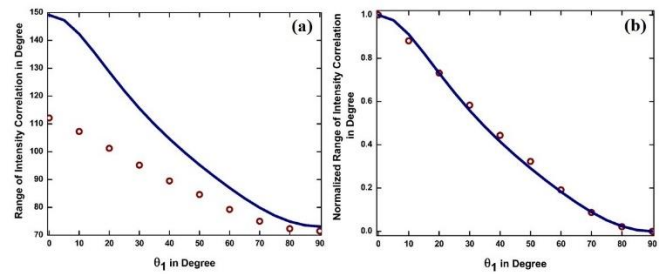

Fig. 4. The experimentally observed (circles) and theoretically predicted (solid line) variations of the (a) range and (b) normalized range of the intensity correlation with $\theta_1$.

It is shown theoretically that the range of the intensity correlation is same for all the polarization components of a completely depolarized random field. In the obtained experimental results, it is observed that the range of the intensity correlation is different for different polarization components, which may be attributed due to the difference in the average intensity of different polarization components of the field. This is confirmed by measuring the average intensities of the horizontal and vertical polarization components of the depolarized field, and the ratio of the average intensity is found to be 1.94 i.e. α = 1.94 in Eq. 10. The value of α is inserted in Eq. 10 and the range of the intensity correlation of different polarization components is calculated theoretically. The predicted variation (solid line) of the range of the intensity correlation with $\theta_1$ along with the experimental results (circles) are presented in Fig. 4(a). The deviation of the experimental results from the theoretical prediction is observed may be due to the fact that although the speckles with different polarizations have different intensity, they are captured at the same exposure time. Although the experimental observations are not matching exactly with the theoretical prediction, in order to compare the nature of these variations, the experimentally observed variation and the theoretical prediction are normalized, and the normalized variations are shown in Fig. 4(b). It can be observed from Fig. 4(b) that the normalized variation of the experimental results are matching well with that of the theoretical prediction, which confirms that although the experimental results are not exactly matching with the theoretical prediction, the nature of these variations are same.

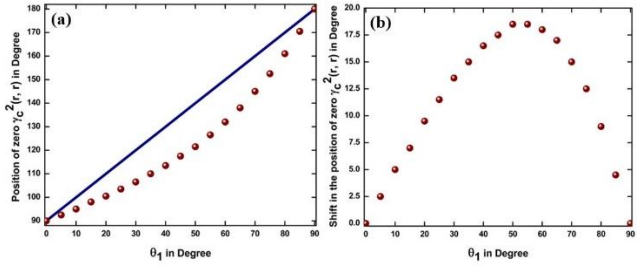

Fig. 5. The (a) variation of the position of zero intensity cross-correlation and (b) shift in the position of zero cross-correlation from the expected value as a function of $\theta_1$.

It can also be observed from Eq. 10 that for non-unit value of $\alpha$, the maximum $\gamma_c^2$ is not zero at $\theta_2 = \theta_1 \pm 90^0$, except for $\theta_1 = 0^0$ and $90^0$. This indicates that except for $\theta_1 = 0^0$ and $90^0$, the cross-correlation of two speckle patterns, filtered from depolarized speckles, with mutual orthogonal polarizations is not zero, if the random field is not completely depolarized. The value of $\theta_2$ at which the maximum $\gamma_c^2$ is zero, denoted as $\theta_{2min}$, is determined theoretically from Eq. 10, and its deviation from $\theta_2 = \theta_1 \pm 90^0$, denoted as $\theta_{2act}$, is calculated for each value of $\theta_1$, which ranges between $0^0$ and $90^0$. The change of $\theta_{2min}$ (circles) along with $\theta_{2act}$ (solid line) with the change of $\theta_1$ are shown in Fig. 5(a), where the deviation of $\theta_{2min}$ from $\theta_{2act}$, except for the case of $\theta_1 = 0^0$ and $90^0$, is clearly visible. The change in the deviation of $\theta_{2min}$ from $\theta_{2act}$ with the change of $\theta_1$ is studied and the variation is shown in Fig. 5(b), where it can be observed that with the increase of $\theta_1$, initially, the deviation increases and reaches to a maximum value and then, starts to decrease. The observed maximum deviation of $\theta_{2min}$ from $\theta_{2act}$ depends on the value of $\alpha$. If $\alpha = 1$, the maximum deviation is zero, and in the present case, a maximum deviation of $18.5^0$ is observed due to $\alpha$ value of 1.94.

In this work, we have investigated the correlation between different polarization components of polarization speckle both theoretically and experimentally following a newly proposed technique referred as the polarization based intensity correlation. We have studied the range of the correlation of different polarization components of the random field, and have observed that if the field is not completely depolarized, the range is different for different polarization components, and the polarization component with higher average intensity has longer range of the intensity correlation. It is also observed that the incomplete depolarization of the speckles leads to non-zero correlation between the orthogonal polarization components of the random field, except for the case of the horizontal and vertical polarization components. This study will be useful for complete characterization of a depolarizing medium, which is necessary for polarization retrieval related study through a bulk scattering medium. The existence of non-zero correlation between the orthogonal polarization components of the polarization speckle may also be useful in speckle cryptography.

**Funding.** Science and Engineering Research Board (SERB), Government of India (SR/FTP/PS-170/2013); Indian Institute of Technology Kharagpur (IIT/SRIC/PHY/VBC/2014-15/43).